\pdfoutput=1
\documentclass[12pt,a4paper]{article}

\usepackage{enumitem}
\usepackage{ifthen} 
\newboolean{pdflatex}
\setboolean{pdflatex}{true} 

\usepackage{placeins}

\newboolean{articletitles}
\setboolean{articletitles}{true} 

\newboolean{uprightparticles}
\setboolean{uprightparticles}{false} 

\usepackage{enumitem}
\setlist[itemize]{noitemsep}
\usepackage{bashful}

\def\paperauthors{LHCb collaboration} 
\def\paperasciititle{Comprehensive analysis of the B0->Kstmumu decay} 
\def\papertitle{Comprehensive analysis of the $\Bz\to\Kstarz\mumu$ decay} 
\def\paperkeywords{{High Energy Physics}, {LHCb}} 
\def\papercopyright{\the\year\ CERN for the benefit of the LHCb collaboration} 
\def\paperlicence{CC BY 4.0 licence}
\def\paperlicenceurl{https://creativecommons.org/licenses/by/4.0/}

\def\mkpi {\ensuremath{m(\Kp\pim)}\xspace}

\def\flavio     {\mbox{\textsc{Flavio}}\xspace}
\def\BF {dBF/d\qq}

\def\Mkpimm{\ensuremath{m(\Kp\pim\mumu)}\xspace} 

\def\BdToKpiJpsi {\decay{\Bz}{\Kp\pim\jpsi}}
\def\BdToKpimm    {\decay{\Bd}{\Kp\pim\mumu}}

\def\BdPsitwosKst  {\decay{\Bd}{\psitwos\Kstarz}}

\def\BzToKstphimm{\decay{\Bz}{\Pphi(1020)(\to\mumu)\Kstarz}}

\def\BsTophiKKmm  {\decay{\Bs}{\Pphi(1020)(\to\Kp\Km)\mumu}}
\def\LambdapKmm    {\decay{\Lb}{\proton\Km\mumu}}

\def\GammaorGammabar {\kern \thebaroffset\optbar{\kern -\thebaroffset \Gamma}\xspace}

\def\AFB {\ensuremath{A_{\rm{FB}}}\xspace}

\def\thetal {\ensuremath{\theta_{\ell}}\xspace}
\def\thetak {\ensuremath{\theta_{K}}\xspace}

\def\BuJpsiK       {\decay{\Bp}{ \jpsi \Kp}}

\def\BuKMuMu     {\decay{\Bp}{ \Kp \mup\mun}}

\newcommand{\comment}[1]{}

\usepackage[top=1in, bottom=1.25in, left=1in, right=1in]{geometry}

\usepackage{microtype}
\usepackage{lineno}  
\usepackage{xspace} 
\usepackage{caption} 

\usepackage{graphicx}  
\usepackage{color}
\usepackage{colortbl}
\graphicspath{{./figs/}} 

\usepackage{amsmath} 
\usepackage{amssymb}
\usepackage{amsfonts}
\usepackage{upgreek} 
\usepackage{booktabs}
\newcommand*\patchAmsMathEnvironmentForLineno[1]{%
\expandafter\let\csname old#1\expandafter\endcsname\csname #1\endcsname
\expandafter\let\csname oldend#1\expandafter\endcsname\csname
end#1\endcsname
 \renewenvironment{#1}%
   {\linenomath\csname old#1\endcsname}%
   {\csname oldend#1\endcsname\endlinenomath}%
}
\newcommand*\patchBothAmsMathEnvironmentsForLineno[1]{%
  \patchAmsMathEnvironmentForLineno{#1}%
  \patchAmsMathEnvironmentForLineno{#1*}%
}
\AtBeginDocument{%
\patchBothAmsMathEnvironmentsForLineno{equation}%
\patchBothAmsMathEnvironmentsForLineno{align}%
\patchBothAmsMathEnvironmentsForLineno{flalign}%
\patchBothAmsMathEnvironmentsForLineno{alignat}%
\patchBothAmsMathEnvironmentsForLineno{gather}%
\patchBothAmsMathEnvironmentsForLineno{multline}%
\patchBothAmsMathEnvironmentsForLineno{eqnarray}%
}

\usepackage[pdftex,
            pdfauthor={\paperauthors},
            pdftitle={\paperasciititle},
            pdfkeywords={\paperkeywords}]{hyperref}
\usepackage{hyperxmp}
\hypersetup{
    pdfcopyright={Copyright (C) \papercopyright},
    pdflicenseurl={\paperlicenceurl}
}

\usepackage[colorinlistoftodos,textsize=scriptsize]{todonotes}

\usepackage[bottom,flushmargin,hang,multiple]{footmisc}

\usepackage[all]{hypcap} 

\usepackage{xspace} 
\usepackage{upgreek}

\def\lhcb   {\mbox{LHCb}\xspace}

\def\MagUp {\mbox{\em Mag\kern -0.05em Up}\xspace}

\ifthenelse{\boolean{uprightparticles}}%
{

 \def\Pmu         {\ensuremath{\upmu}\xspace}

 \def\Ppi         {\ensuremath{\uppi}\xspace}

 \def\Pphi        {\ensuremath{\upphi}\xspace}

 \def\Ppsi        {\ensuremath{\uppsi}\xspace}

 \def\PDelta      {\ensuremath{\Delta}\xspace}                 
 \def\PXi         {\ensuremath{\Xi}\xspace}                 
 \def\PLambda     {\ensuremath{\Lambda}\xspace}                 
 \def\PSigma      {\ensuremath{\Sigma}\xspace}                 
 \def\POmega      {\ensuremath{\Omega}\xspace}                 
 \def\PUpsilon    {\ensuremath{\Upsilon}\xspace}
 \let\oldPi\Pi
 \def\PPi         {\ensuremath{\oldPi}\xspace}

 \def\PB      {\ensuremath{\mathrm{B}}\xspace}                 
                  
 \def\PD      {\ensuremath{\mathrm{D}}\xspace}

 \def\PJ      {\ensuremath{\mathrm{J}}\xspace}                 
 \def\PK      {\ensuremath{\mathrm{K}}\xspace}

 \def\Pb      {\ensuremath{\mathrm{b}}\xspace}                 
 \def\Pc      {\ensuremath{\mathrm{c}}\xspace}

 \def\Pi      {\ensuremath{\mathrm{i}}\xspace}

 \def\Pp      {\ensuremath{\mathrm{p}}\xspace}

 \def\Ps      {\ensuremath{\mathrm{s}}\xspace}

 \def\thebaroffset{0.0em}
}
{

 \def\Pmu         {\ensuremath{\mu}\xspace}

 \def\Ppi         {\ensuremath{\pi}\xspace}

 \def\Pphi        {\ensuremath{\phi}\xspace}

 \def\Ppsi        {\ensuremath{\psi}\xspace}                 
                  
 \mathchardef\PDelta="7101
 \mathchardef\PXi="7104
 \mathchardef\PLambda="7103
 \mathchardef\PSigma="7106
 \mathchardef\POmega="710A
 \mathchardef\PUpsilon="7107
 \mathchardef\PPi="7105
                  
 \def\PB      {\ensuremath{B}\xspace}                 
                  
 \def\PD      {\ensuremath{D}\xspace}

 \def\PJ      {\ensuremath{J}\xspace}                 
 \def\PK      {\ensuremath{K}\xspace}

 \def\Pb      {\ensuremath{b}\xspace}                 
 \def\Pc      {\ensuremath{c}\xspace}

 \def\Pi      {\ensuremath{i}\xspace}

 \def\Pp      {\ensuremath{p}\xspace}

 \def\Ps      {\ensuremath{s}\xspace}

 \def\thebaroffset{0.18em}
}
\newcommand{\offsetoverline}[2][\thebaroffset]{\kern #1\overline{\kern -#1 #2}}%

\makeatletter
\ifcase \@ptsize \relax
  \newcommand{\miniscule}{\@setfontsize\miniscule{4}{5}}
\or
  \newcommand{\miniscule}{\@setfontsize\miniscule{5}{6}}
\or
  \newcommand{\miniscule}{\@setfontsize\miniscule{5}{6}}
\fi
\makeatother

\DeclareRobustCommand{\optbar}[1]{\shortstack{{\miniscule (\rule[.5ex]{1.25em}{.18mm})}
  \\ [-.7ex] $#1$}}

\def\mup        {{\ensuremath{\Pmu^+}}\xspace}
\def\mun        {{\ensuremath{\Pmu^-}}\xspace} 

\def\mumu       {{\ensuremath{\Pmu^+\Pmu^-}}\xspace}

\def\squark    {{\ensuremath{\Ps}}\xspace}

\def\cquark    {{\ensuremath{\Pc}}\xspace}
\def\cquarkbar {{\ensuremath{\overline \cquark}}\xspace}
\def\ccbar     {{\ensuremath{\cquark\cquarkbar}}\xspace}
\def\bquark    {{\ensuremath{\Pb}}\xspace}

\def\pion   {{\ensuremath{\Ppi}}\xspace}

\def\pim    {{\ensuremath{\pion^-}}\xspace}

\def\kaon    {{\ensuremath{\PK}}\xspace}
\def\Kbar    {{\ensuremath{\offsetoverline{\PK}}}\xspace}

\def\KorKbar {\kern \thebaroffset\optbar{\kern -\thebaroffset \PK}{}\xspace}

\def\Kp      {{\ensuremath{\kaon^+}}\xspace}
\def\Km      {{\ensuremath{\kaon^-}}\xspace}

\def\Kstarz  {{\ensuremath{\kaon^{*0}}}\xspace}
\def\Kstarzb {{\ensuremath{\Kbar{}^{*0}}}\xspace}
\def\Kstar   {{\ensuremath{\kaon^*}}\xspace}

\def\Dbar    {{\ensuremath{\offsetoverline{\PD}}}\xspace}
\def\D       {{\ensuremath{\PD}}\xspace}

\def\DorDbar {\kern \thebaroffset\optbar{\kern -\thebaroffset \PD}\xspace}

\def\Dp      {{\ensuremath{\D^+}}\xspace}
\def\Dm      {{\ensuremath{\D^-}}\xspace}

\def\DpDm    {\ensuremath{\Dp {\kern -0.16em \Dm}}\xspace}

\def\B       {{\ensuremath{\PB}}\xspace}
\def\Bbar    {{\ensuremath{\offsetoverline{\PB}}}\xspace}

\def\BorBbar {\kern \thebaroffset\optbar{\kern -\thebaroffset \PB}\xspace}
\def\Bz      {{\ensuremath{\B^0}}\xspace}
\def\Bzb     {{\ensuremath{\Bbar{}^0}}\xspace}
\def\Bd      {{\ensuremath{\B^0}}\xspace}
\def\Bdb     {{\ensuremath{\Bbar{}^0}}\xspace}
\def\BdorBdbar {\kern \thebaroffset\optbar{\kern -\thebaroffset \Bd}\xspace}
\def\Bu      {{\ensuremath{\B^+}}\xspace}

\def\Bp      {{\ensuremath{\Bu}}\xspace}

\def\Bs      {{\ensuremath{\B^0_\squark}}\xspace}
\def\Bsb     {{\ensuremath{\Bbar{}^0_\squark}}\xspace}
\def\BsorBsbar {\kern \thebaroffset\optbar{\kern -\thebaroffset \Bs}\xspace}

\def\jpsi     {{\ensuremath{{\PJ\mskip -3mu/\mskip -2mu\Ppsi}}}\xspace}
\def\psitwos  {{\ensuremath{\Ppsi{(2S)}}}\xspace}

\def\Y#1S{\ensuremath{\PUpsilon{(#1S)}}\xspace}

\def\proton      {{\ensuremath{\Pp}}\xspace}

\def\Lz          {{\ensuremath{\PLambda}}\xspace}

\def\LorLbar     {\kern \thebaroffset\optbar{\kern -\thebaroffset \PLambda}\xspace}

\def\Lb           {{\ensuremath{\Lz^0_\bquark}}\xspace}

\def\BF         {{\ensuremath{\mathcal{B}}}\xspace}

\newcommand{\decay}[2]{\ensuremath{#1\!\to #2}\xspace} 

\def\to                 {\ensuremath{\rightarrow}\xspace}

\def\qsq       {{\ensuremath{q^2}}\xspace}

\def\eps   {{\ensuremath{\varepsilon}}\xspace}

\def\CP                {{\ensuremath{C\!P}}\xspace}

\def\BdToKstmm    {\decay{\Bd}{\Kstarz\mup\mun}}

\def\BdToKpimm    {\decay{\Bd}{\Kp\pim\mup\mun}}

\def\BdToJPsiKst  {\decay{\Bd}{\jpsi\Kstarz}}

\def\BuJpsiK      {\decay{\Bu}{\jpsi\Kp}}

\def\AFB      {\ensuremath{A_{\mathrm{FB}}}\xspace}

\def\AT#1     {\ensuremath{A_{\mathrm{T}}^{#1}}\xspace}           

\def\C#1      {\ensuremath{\mathcal{C}_{#1}}\xspace}                       
\def\Cp#1     {\ensuremath{\mathcal{C}_{#1}^{'}}\xspace}                    
\def\Ceff#1   {\ensuremath{\mathcal{C}_{#1}^{\mathrm{(eff)}}}\xspace}        
\def\Cpeff#1  {\ensuremath{\mathcal{C}_{#1}^{'\mathrm{(eff)}}}\xspace}       
\def\Ope#1    {\ensuremath{\mathcal{O}_{#1}}\xspace}                       
\def\Opep#1   {\ensuremath{\mathcal{O}_{#1}^{'}}\xspace}                    

\newcommand{\aunit}[1]{\ensuremath{\text{\,#1}}}       

\newcommand{\tev}{\aunit{Te\kern -0.1em V}\xspace}
\newcommand{\gev}{\aunit{Ge\kern -0.1em V}\xspace}
\newcommand{\mev}{\aunit{Me\kern -0.1em V}\xspace}
\newcommand{\kev}{\aunit{ke\kern -0.1em V}\xspace}
\newcommand{\ev}{\aunit{e\kern -0.1em V}\xspace}
 
\newcommand{\mevc}{\ensuremath{\aunit{Me\kern -0.1em V\!/}c}\xspace}
\newcommand{\gevc}{\ensuremath{\aunit{Ge\kern -0.1em V\!/}c}\xspace}
\newcommand{\mevcc}{\ensuremath{\aunit{Me\kern -0.1em V\!/}c^2}\xspace}
\newcommand{\gevcc}{\ensuremath{\aunit{Ge\kern -0.1em V\!/}c^2}\xspace}
\newcommand{\gevgevcccc}{\ensuremath{\gev^2\!/c^4}\xspace} 

\def\fb   {\ensuremath{\aunit{fb}}\xspace}
\def\invfb   {\ensuremath{\fb^{-1}}\xspace}

\def\deriv {\ensuremath{\mathrm{d}}}

\def\gsim{{~\raise.15em\hbox{$>$}\kern-.85em
          \lower.35em\hbox{$\sim$}~}\xspace}
\def\lsim{{~\raise.15em\hbox{$<$}\kern-.85em
          \lower.35em\hbox{$\sim$}~}\xspace}

\newcommand{\Real}{\ensuremath{\mathcal{R}e}\xspace}
\newcommand{\Imag}{\ensuremath{\mathcal{I}m}\xspace}

\def\sPlot{\mbox{\em sPlot}\xspace}

\def\tell1  {TELL1\xspace}
\def\ukl1   {UKL1\xspace}

\newcommand{\eg}{\mbox{\itshape e.g.}\xspace}

\newcommand{\lhcborcid}[1]{\href{https://orcid.org/#1}{\hspace*{0.1em}\raisebox{-0.45ex}{\includegraphics[width=1em]{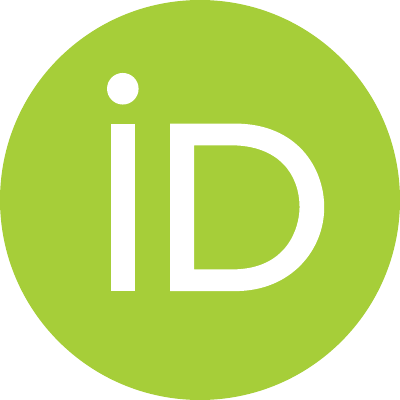}}}}

\usepackage{cite} 
\usepackage{mciteplus}

\usepackage{longtable} 
\usepackage{bm}
\begin{document}

\renewcommand{\thefootnote}{\fnsymbol{footnote}}
\setcounter{footnote}{1}


\begin{titlepage}
\pagenumbering{roman}

\vspace*{-1.5cm}
\centerline{\large EUROPEAN ORGANIZATION FOR NUCLEAR RESEARCH (CERN)}
\vspace*{1.5cm}
\noindent
\begin{tabular*}{\linewidth}{lc@{\extracolsep{\fill}}r@{\extracolsep{0pt}}}
\ifthenelse{\boolean{pdflatex}}
{\vspace*{-1.5cm}\mbox{\!\!\!\includegraphics[width=.14\textwidth]{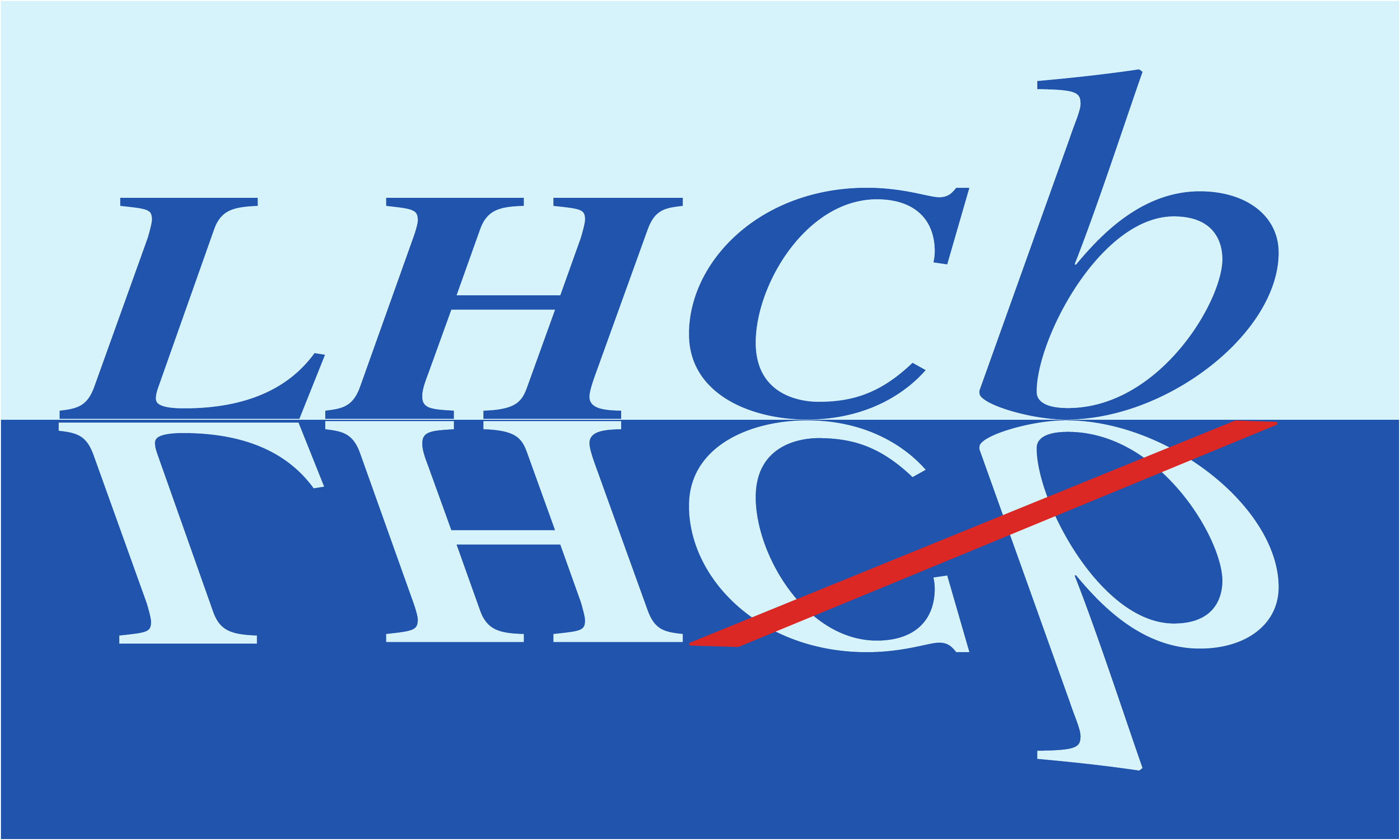}} & &}%
{\vspace*{-1.2cm}\mbox{\!\!\!\includegraphics[width=.12\textwidth]{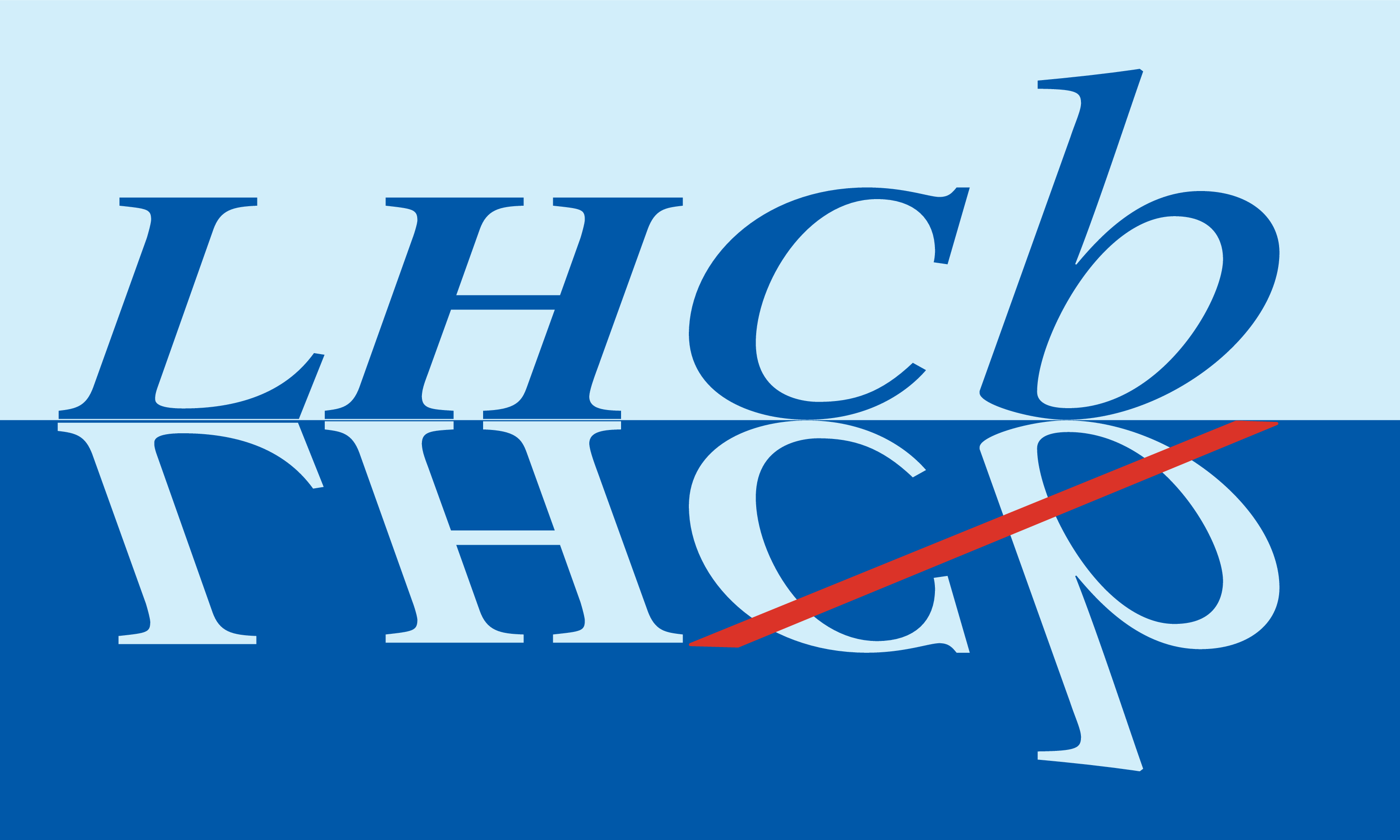}} & &}%
\\
 & & CERN-EP-2025-278 \\  
 & & LHCb-PAPER-2025-041  \\  
 & & 8 July 2026\\ 
 & & \\
\end{tabular*}

\vspace*{4.0cm}

{\normalfont\bfseries\boldmath\huge
\begin{center}
  \papertitle 
\end{center}
}

\vspace*{2.0cm}

\begin{center}
\paperauthors\footnote{Authors are listed at the end of this paper.}
\end{center}

\vspace{\fill}

\begin{abstract}
  \noindent
An analysis of the $B^{0}\rightarrow K^{*0}(\to\Kp\pim)\mu^{+}\mu^{-}$ decay is presented using proton-proton collision data collected by the \lhcb experiment, corresponding to an integrated luminosity of $8.4\invfb$.
The full set of \CP-averaged and \CP-asymmetric angular observables is determined in bins of the invariant mass squared of the dimuon system, as well as the branching fraction relative to the $B^{0}\rightarrow  \jpsi(\to\mu^{+}\mu^{-})\Kp\pim$ decay. For the first time, the full set of observables pertaining to the $\Kp\pim$ S-wave contribution to the final state are presented and consideration is given to effects arising from the mass of the muons. The extracted \CP-asymmetry observables show no significant deviations from zero. The measurements of the \CP-averaged observables and the branching fractions continue to exhibit the pattern of tensions with the Standard Model predictions that have been seen in previous analyses that use part of the dataset considered in this Letter.

\end{abstract}

\vspace*{2.0cm}

\begin{center}
  Published in Physical Review Letters 137 (2026) 021802. 
\end{center}

\vspace{\fill}

{\footnotesize 
\centerline{\copyright~\papercopyright. \href{\paperlicenceurl}{\paperlicence}.}}
\vspace*{2mm}

\end{titlepage}


\newpage
\setcounter{page}{2}
\mbox{~}


\renewcommand{\thefootnote}{\arabic{footnote}}
\setcounter{footnote}{0}

\cleardoublepage


\pagestyle{plain} 
\setcounter{page}{1}
\pagenumbering{arabic}


An intriguing set of discrepancies between experimental measurements and the predictions of the Standard Model (SM) has been observed in the $B^{0}\rightarrow K^{*0}\mu^{+}\mu^{-}$ decay~\cite{LHCb-PAPER-2015-051,LHCb-PAPER-2016-012,Belle:2016fev,ATLAS:2018gqc,LHCb-PAPER-2020-002,CMS:2024atz} and in several other processes mediated by the same underlying \decay{b}{s \mup\mun} quark transition~\cite{LHCb-PAPER-2014-006,LHCb-PAPER-2020-041,LHCb-PAPER-2021-014,LHCb-PAPER-2021-022}.
In the SM, these flavor-changing neutral currents occur only via virtual quantum loops. The resulting suppression means that as yet undiscovered particles, which could have masses that make them inaccessible to direct searches, may contribute to these decays at a level comparable to the SM processes. For example, tensions with SM predictions may be explained by New Physics models that introduce an additional contribution to the effective four-fermion $\bquark\squark\mu\mu$ vector coupling, \C9 \cite{Descotes-Genon:2013wba,Beaujean:2013soa,Altmannshofer:2014rta, Capdevila:2017bsm,London:2021lfn,Gubernari:2022hxn,Greljo:2022jac,Ciuchini:2022wbq,Alguero:2023jeh}.
However, despite numerous theoretical developments~\cite{ Khodjamirian:2010vf, Khodjamirian:2012rm,Bouchard:2013eph,Horgan:2013hoa,Lyon:2014hpa,Horgan:2015vla,Bailey:2015dka,Bharucha:2015bzk,Agadjanov:2016fbd,Khodjamirian:2017fxg, Bobeth:2017vxj,Gubernari:2018wyi, Descotes-Genon:2019bud,Gao:2019lta,Gubernari:2020eft,Gubernari:2023puw, Descotes-Genon:2023ukb,PhysRevD.107.014511}, there is still debate about whether the observations might instead be explained by SM long-distance effects~\cite{Khodjamirian:2010vf,Korchin:2012kz,Jager:2012uw, Descotes-Genon:2013wba,Lyon:2014hpa,Jager:2014rwa,Ciuchini:2015qxb,Gubernari:2020eft,Ciuchini:2022wbq,Gubernari:2022hxn,PhysRevD.111.093007,isidori2025charmrescatteringb0tok0barellell}. Recent LHCb analyses that try to quantify these effects by fitting \BdToKstmm data together with ancillary data and theoretical predictions do not suggest substantially enhanced long-distance effects~\cite{LHCb-PAPER-2023-033, LHCb-PAPER-2024-011}. 

This Letter presents the most precise measurements to date of the \CP-averaged angular observables of the \BdToKstmm decay, the associated \CP asymmetries and the differential branching fraction. The $\Kstarz$ resonance, where \Kstarz denotes the $\Kstar(892)^0$ meson, is reconstructed through its decay to $\Kp\pim$. This final state receives contributions from decays with the \Kp\pim in an S-wave configuration, as well as the P-wave configuration from the \Kstarz.
The reported observables correspond to averages over different regions of the dimuon invariant mass squared, \qsq.
The dataset used corresponds to an integrated luminosity of $8.4\invfb$ of proton-proton collisions collected by the \lhcb experiment during 2011, 2012, and 2016–2018. Throughout this Letter charge-conjugate processes are implied unless otherwise stated.

The analysis broadly follows the strategy of previous \qsq-binned studies of this decay mode, using LHCb data collected during 2011–2016~\cite{LHCb-PAPER-2011-020, LHCb-PAPER-2013-019, LHCb-PAPER-2013-037, LHCb-PAPER-2014-032, LHCb-PAPER-2015-051, LHCb-PAPER-2016-012, LHCb-PAPER-2020-002}. The present study introduces several significant innovations over those earlier analyses. These include: an updated expression for interference effects in the differential decay rate; the first simultaneous treatment of all relevant kinematic variables for describing the decay, incorporating for the first time the invariant mass of the kaon-pion system, $m(\Kp\pim)$, which is also studied in a wider range than previously considered; the first treatment of the effect of lepton masses; the first simultaneous measurement of the differential branching fraction alongside the angular observables; and the first measurement of the full set of \CP asymmetries. The inclusion of $\mkpi$ directly when measuring the angular distributions gives sensitivity to new observables related to the interference between P- and S-wave contributions~\cite{Alguero:2021yus}.
In this study effects associated with lepton masses are found to have an important impact across a range of \qsq values below the charmonium resonances.
The approach taken for the simultaneous measurement of the differential branching fraction accounts for any variation of the angular decay rate due to potential New Physics contributions, 
while the first measurement of the full set of \CP asymmetries is provided alongside the full covariance matrix relating the \CP-averaged and \CP-asymmetric observables. Narrower \qsq bins compared to those from previous analyses are included to gain an improved understanding of the \qsq dependence of the measured observables. The offline selection has also been reoptimized compared to the previous analysis~\cite{LHCb-PAPER-2020-002}. 
Finally, the analysis described in this Letter was performed twice, using two independently developed frameworks, providing an important validation of the results. 

The \BdToKpimm decay is completely described by five kinematic variables: \qsq, \mkpi, and the three decay angles, $\vec{\Omega}=(\cos\thetal,\cos\thetak,\phi)$ defined in Ref.~\cite{LHCb-PAPER-2013-019}. 
The resulting differential decay rate is parameterized by approximately 20 \CP-averaged angular observables, with a similar number describing the difference in angular distributions between \Bz and \Bzb decays (\CP asymmetries). These observables are determined for candidates with \mbox{$0.7459 < \mkpi < 1.0959\gevcc$}, except for the \mbox{$17 < \qsq < 19\gevgevcccc$} region, where a narrower \mbox{$0.7459 < \mkpi < 0.9590\gevcc$} range is applied.
This differential rate is described in Ref.~\cite{Alguero:2021yus}, with the specific form used for this analysis provided in the End Matter.
Observables accessible in the \BdToKstmm decay have previously been extensively studied~\cite{Ali:1999mm, Kruger:1999xa, Kruger:2005ep, Altmannshofer:2008dz,  Egede:2008uy, Becirevic:2011bp, Bobeth:2010wg, Egede:2010zc, Bobeth:2011gi, Bobeth:2012vn, Matias:2012xw, Matias:2012qz, Bobeth:2008ij, Descotes-Genon:2012isb, Kumar:2014bna}.
The analysis presented here measures: the \CP-averaged and \CP-asymmetry observables, $S_{i}$ and $A_{i}$, for both P-wave and S-wave, including the interference between the two; the P-wave decay rate; and the $P_{i}^{(\prime)}$ observables that are constructed to cancel the leading-order theoretical uncertainties from the \decay{\Bz}{\Kstarz} form factors~\cite{Descotes-Genon:2012isb}.

As detailed in the End Matter, the total basis of observables may be reduced by making assumptions on the effect of the lepton mass, or the presence of New Physics scalar or tensor amplitudes. Without these assumptions, 
significant correlations between some observables can lead to sizeable biases or increased uncertainties in the measured quantities, meaning that a reduced model gives better precision on the extracted observables. Given the precision of this analysis, the effect of the muon mass is found to be significant up to $\qsq\approx 6\gevgevcccc$, with the most pronounced effect at the lowest \qsq values. Therefore, for the first time, this analysis accounts for the lepton mass when determining the angular observables.

The LHCb detector is a single-arm forward spectrometer covering the pseudorapidity range $2< \eta < 5$, described in detail in Refs.~\cite{LHCb-DP-2008-001, LHCb-DP-2014-002}. The polarity of the LHCb dipole magnet is periodically reversed to cancel the effects of oppositely charged particles being deflected in opposite directions. The online event selection is performed by a trigger, which comprises a hardware stage, based on information from the calorimeter and muon systems, followed by a software stage~\cite{LHCb-DP-2012-004}. Data selected by the trigger undergo a further, centralized, offline processing step~\cite{Stripping}.

Signal candidates are formed from a pair of oppositely charged tracks that are identified as muons, combined with a \Kstarz candidate that is formed from tracks identified as a kaon and a pion.  The tracks of the final-state particles are required to have significant impact parameter with respect to all primary proton-proton interaction vertices~(PVs) in the event and to form a common vertex of good quality. 
The \Bz-candidate decay vertex is required to be significantly displaced from all the PVs. 
It is also required that the angle between the reconstructed \Bz momentum and the vector connecting one of the PVs to the reconstructed \Bz decay vertex, $\theta_{\rm DIRA}$, is small.

Large samples of simulated events, produced with the software described in Refs.~\cite{Sjostrand:2007gs, Lange:2001uf, Allison:2006ve, *Agostinelli:2002hh}, are used to assess the reconstruction and selection efficiencies for signal candidates and to estimate the contamination from residual backgrounds. Inaccuracies in the simulation are corrected using control samples from data. The simulated particle-identification (PID) response is corrected using dedicated data control samples~\cite{LHCb-PUB-2016-021,LHCb-DP-2018-001}, as is the track-reconstruction efficiency~\cite{LHCb-DP-2013-002}. 
Data-simulation differences arising from mismodeled $\Bz$ kinematics, event multiplicity, hardware trigger efficiency, track and vertex quality are corrected using two separate control modes; $\BdToJPsiKst$ and $\BuJpsiK$.
The results from the two modes are compared and found to be in good agreement, with the differences in the corrections used to evaluate a systematic uncertainty.

Residual backgrounds consist of a combinatorial component, formed from fragments of multiple decays, with distributions that vary smoothly in \Mkpimm, \mkpi and $\vec{\Omega}$; and peaking backgrounds, where a single decay is selected but with some of the final-state particles misidentified or misreconstructed, which tend to accumulate in specific \Mkpimm and $\vec{\Omega}$ regions. The combinatorial background is reduced by a dedicated boosted decision tree~(BDT) classifier~\cite{Breiman,AdaBoost}. The classifier is trained using \BdToJPsiKst candidates in the $8.0<\qsq<11.0\gevgevcccc$ region as a signal proxy, where the background contribution has been statistically subtracted using the \sPlot procedure~\cite{Pivk:2004ty}. For the background training sample, candidates with invariant mass $5.5<\Mkpimm<5.7\gevcc$ are used. Separate BDTs are trained using a cross-validation technique~\cite{Blum:1999} for each of the three run periods (2011--2012, 2016, 2017--2018), which reflect significant changes in the data-taking conditions.
The variables used as inputs to the BDT classifier include the reconstructed \Bz decay time and the vertex-fit quality, as well as its momentum, transverse momentum, and $\theta_{\rm DIRA}$. Additionally, PID information from the Ring Imaging Cherenkov detectors and the muon system, and variables describing the isolation of the final-state tracks~\cite{LHCb-PAPER-2011-004}, are used. 
A requirement is placed on the BDT output to maximize the significance of $\BdToKstmm$ signal decays. This requirement rejects more than $98\%$ of the remaining combinatorial background, while retaining more than $89\%$ of the signal.

Dedicated vetoes are applied to reduce peaking backgrounds from processes such as \BsTophiKKmm, \LambdapKmm, \BdToJPsiKst, \BdPsitwosKst and \decay{\Bdb}{\Kstarzb\mup\mun}, where at least one final-state particle has been misidentified. The last decay forms a background if both the kaon and pion are misidentified. The decays involving charmonia can mimic the signal decay if a muon is misidentified as a hadron and vice versa. The vetoes use the invariant mass of the candidates recomputed with appropriate changes in the particle mass hypotheses, and the output of a neural network based on variables that combine PID and track information about the particle~\cite{LHCb-PUB-2016-021}. 
The \BuKMuMu decay can constitute a background if the final-state particles are combined with a random pion produced by other processes in the event. This background is vetoed by removing candidates with $\Mkpimm>5.38\gevcc$ if they have a $\Kp\mumu$ invariant mass within $\pm 0.06\gevcc$ of the known \Bp mass~\cite{PDG2024}.

The analogous background due to \BuJpsiK decays remains significant for \mbox{$\qsq \gtrsim 6\gevgevcccc$}, owing to the long radiative tail of the \jpsi resonance and its large branching fraction to the dimuon final state. As a result of this radiative tail, these candidates have a $m({\Kp\mumu})$ value below the known \Bp mass and so escape the veto for the \BuKMuMu background. A dedicated BDT classifier, trained using simulation to represent both the signal and this particular background, is used to suppress these candidates to a negligible level. Similarly, a background can occur from cascades of semileptonic decays with large branching fractions, \eg \decay{B}{\Dbar(\to \Kp\mu^{-}X)\mu^{+}X}, when incorrectly combined with a pion selected from elsewhere in the event. Here, $X$ can represent any additional unreconstructed particles in the final state. Due to the inherent asymmetry of the momenta of the two muons, these decays lead to a characteristic asymmetry in $\cos\theta_{\ell}$. 
Another dedicated BDT classifier is trained to reduce these cascade backgrounds to negligible levels. For the training of this BDT, data candidates with $\Mkpimm>5.35\gevcc$, enriched in semileptonic contributions by requiring $m(\Kp\mun)<1.9 \gevcc $ and $\qsq<8.0\gevgevcccc$, serve as a background proxy, while simulated candidates with $m(\Kp\mun)<1.9 \gevcc $ are used as a signal proxy. The most discriminating variables include the difference in three-momenta between the $\Kstarz$ candidate and dimuon pair and the transverse momentum of the pion.
The Cabibbo-suppressed background from \decay{\Bsb}{\Kstarz\mup\mun} decays is expected to be at the level of 1\% of the signal.
The analysis is performed with and without this background component included in the fit, and the effect on the measured decay rate is found to be negligible.
Backgrounds from \bquark-hadron decays to fully hadronic final states, where two hadrons are misidentified as muons, are negligible. 
The residual contamination from peaking backgrounds is estimated to be at the level of 2\% of the signal. Such backgrounds are not modeled in the analysis procedure and are considered only as a source of systematic uncertainty.

Candidates are required to have \mbox{$5.17<\Mkpimm<5.70\gevcc$} and \mbox{$0.7459<\mkpi<1.0959\gevcc$}. The larger \mkpi window used here compared to previous analyses increases the yield of S-wave signal candidates, which allows for a more precise determination of the S-wave fraction, $F_{\rm S}$, and the observables pertaining to the interference of P- and S-wave amplitudes.
Backgrounds from the decays \decay{\Bd}{\jpsi(\to\mumu)\Kstarz}, \decay{\Bd}{\psitwos(\to\mumu)\Kstarz} and \mbox{\BzToKstphimm}, which result in the same final state as the signal, are removed by rejecting candidates with \qsq in the ranges \mbox{$8.0 < \qsq < 11.0 \gevgevcccc$}, \mbox{$12.5 < \qsq < 15.0 \gevgevcccc$} and \mbox{$0.98 < \qsq < 1.10 \gevgevcccc$}. The former two decays involve tree-level \decay{\bquark}{\ccbar \squark} transitions and are used to validate several aspects of the analysis procedure.

An extended unbinned maximum-likelihood fit of the differential decay rate as a function of \mkpi, the three decay angles, and \Mkpimm is used to determine the angular observables and differential branching fraction relative to the normalisation mode \BdToKpiJpsi
in each bin of \qsq. This five-dimensional approach introduces new interference observables~\cite{Alguero:2021yus} compared to previous iterations of this analysis and provides the optimal per-event sensitivity to the angular observables. The selected candidates are split into \Bz and \Bzb samples, as well as by the three run periods described above, resulting in six distinct samples that are fit simultaneously, allowing for the full set of \CP-averaged and \CP-asymmetry observables to be extracted in a single fit, with all correlations determined. A model-independent measurement of both the relative branching fractions and the full set of \CP-asymmetry observables is made by adding an extended term to the likelihood, which uses the fitted yields of \Bd and \Bdb events to constrain their respective signal decay rates. This model independence is achieved by calculating the efficiencies of the signal and normalisation modes through integrating the angular distributions of the fit probability density functions (PDFs). 

The distortion of the ($\vec{\Omega}$, \mkpi, \qsq) distribution due to the detector acceptance and the selection is modeled with an acceptance function. This five-dimensional function is parameterized with Legendre polynomials $L$ and coefficients $c$ as
\begin{align}
  \eps(\vec{\Omega},q^2,\mkpi) &= \sum_{ijmno} c_{ijmno} L_i(\cos\thetal) L_j(\cos\thetak) L_m(\phi) L_n(q^2) L_o(\mkpi) \,,
\label{eq:acceptance}
\end{align}
and is computed using simulation. Owing to the rapid variation of the acceptance at low $q^{2}$, which would require high-order polynomials, the acceptance function is split into two regions of \qsq, separated by the $\phi$ resonance at $\qsq\approx 1\gevgevcccc$. In order to account for detection asymmetries, the acceptance function is obtained separately for \Bz and \Bzb decays, and is further split by the three run periods. A goodness-of-fit method using BDTs~\cite{Weisser:2016cnc} is employed to ascertain the minimal polynomial orders needed to properly describe the simulated events for each run period. The resulting acceptance functions are incorporated in the fit, either by directly including them in the PDF from Eq.~\ref{supp:eq:pdf} (see End Matter), or by taking the inverse of the acceptance function as per-event weights.
The latter method is employed when the function varies strongly across the \qsq bin width.

Each of the six subsamples has an extended term in the likelihood, allowing the number of signal events to be determined. 
The yields are normalized to the inclusive \decay{\Bz}{\Kp\pim \jpsi(\to\mumu)} decay mode, with the branching fraction taken from Ref.~\cite{Belle:2014nuw} and scaled by a correction factor derived from the measured amplitudes in Ref.~\cite{Belle:2014nuw} to account for the fraction of decays that fall within the \mkpi window of this analysis. When accounting for the \decay{\jpsi}{\mumu} decay, this results in the normalisation branching fraction ($\mathcal{B}$) being 
\begin{equation}
\BF(\decay{\Bz}{\Kp\pim\jpsi(\to\mumu)} |_{0.7459<\mkpi<1.0959\gevcc}) = (4.88\pm 0.22)\times 10^{-5}. \nonumber
\end{equation}
The uncertainty on this branching fraction partially arises from the uncertainties on the measured amplitudes, which, due to the covariance matrix not being provided in Ref.~\cite{Belle:2014nuw}, are treated as uncorrelated.

The extended term of each sample of the fit calculates the expected signal yield relative to the corresponding sample of the normalisation mode, and so effects from \Bz-\Bzb production and final-state detection asymmetries cancel. The total \CP asymmetry of the \decay{\Bz}{\Kp\pim\jpsi} decay is well-measured and consistent with zero~\cite{LHCb-PAPER-2013-023}.  
The \decay{\Bz}{\Kp\pim\jpsi} yields and efficiencies are determined in dedicated fits for each sample, using the same acceptance and fit functions.

The background distribution is modeled with Chebyshev polynomials of either first, second or fourth order in $\cos\thetal$, $\cos\thetak$, $\phi$ and \mkpi. The coefficients are allowed to vary in the fit and the resulting distribution is assumed to factorize between these variables and \Mkpimm. Furthermore, the shape of the background is assumed to be the same for the \Bz and \Bzb samples. These assumptions are verified using data with $\Mkpimm>5.35\gevcc$.

The \mkpi distribution of the signal is modeled with the amplitudes $\mathcal{BW}_{\rm P}(\mkpi)$ and $\mathcal{L}_{\rm S}(\mkpi)$ for the P- and S-wave components, respectively. The P-wave amplitude includes only the 
$K^{\ast}(892)^0$ resonance and is modeled using a relativistic Breit--Wigner function with mass and width taken from Ref.~\cite{PDG2024} and the standard Blatt--Weisskopf angular momentum factors~\cite{Blatt:1952ije}. Contributions from higher-mass spin-1 resonances are found to be negligible in the analyzed \mkpi region. The amplitude $\mathcal{L}_{\rm S}(\mkpi)$ is taken to be a LASS function~\cite{ASTON1988493,Rui:2017hks} with parameters determined from a fit to the \decay{\Bz}{\Kp\pim\jpsi} decay mode, except for the range parameter of the LASS function, $r$, which is fixed to $1.7\,(\!\gevc)^{-1}$~\cite{LHCb-PAPER-2014-014}. The value of the scattering parameter, which is extracted from the fit to the \decay{\Bz}{\Kp\pim\jpsi} decay mode, is fixed to $3.25\,(\!\gevc)^{-1}$ in the fit to the signal candidates. The barrier radius factor of the decaying \Kstarz meson is fixed to $2.12\,(\!\gevc)^{-1}$, a value again extracted from the fit to the \decay{\Bz}{\Kp\pim\jpsi} decay mode. The analogous radius of the \Bz meson is fixed to $1.6\,(\!\gevc)^{-1}$~\cite{Belle:2013shl}.
The signal shape in \Mkpimm is modeled from simulation using the sum of two Crystal Ball functions~\cite{Skwarnicki:1986xj} with tails to opposite sides, a common peak position, and width scaling factors accounting for differences between data and simulation derived from the \decay{\Bz}{\Kp\pim\jpsi} fits. The combinatorial background is described by an exponential distribution in \Mkpimm.

Several fits of the data are performed with different angular observables floated, corresponding to different assumptions about the underlying physics. Making such assumptions reduces the number of observables, resulting in improved statistical behavior and better precision. Allowing for massive leptons, as well as new physics scalar and tensor amplitudes, results in 25 \CP-averaged angular observables. This is referred to as the “fully massive” model. Due to the large number of correlated parameters, this model generally exhibits reduced statistical precision for the measurement of individual observables.
Unless stated otherwise, the region $\qsq<1.0\gevgevcccc$ is always fitted using the fully massive model.
If leptons are assumed to be massless, two relations between the angular observables can be imposed: $S_{2}^c=- S_1^c$  and $S_1^s=3S_2^s$. This is referred to as the “massless” model. The relation  $S_{2}^c=- S_1^c$ can introduce a bias for lower values of \qsq, as $S_1^c$ has a dependence on the time-like amplitude that is suppressed by a factor of $\frac{4m_{\ell}^{2}}{\qsq}$~\cite{Bobeth:2012vn}. Imposing only the relation $S_1^s=3S_2^s$ avoids this bias and is referred to as the ``partially massive'' model. This is discussed in more detail in the End Matter. 
A fit configuration is defined according to the model, \qsq binning scheme and whether $S_i$ or $P_i^{(\prime)}$ angular observables are extracted. These observables differ in their normalisation, which is more complex for the $P_i^{(\prime)}$ case, see the End Matter and Ref.~\cite{Descotes-Genon:2012isb} for details.  In total, six configurations are implemented:

\begin{enumerate}[label=(\roman*)]
 \item Partially  massive model, $S_i$ basis, with \CP asymmetries fixed to zero;
 \item Partially  massive model, $P_i^{(\prime)}$ basis, with \CP asymmetries fixed to zero;
 \item Fully massive model, $S_i$ basis, with \CP asymmetries fixed to zero;
 \item Massless model, $S_i$ basis, with both \CP-averaged and \CP-asymmetry observables floated;
 \item Massless model, applied also in the $\qsq<1.0\gevgevcccc$ region, $P_i$ basis, with \CP asymmetries fixed to zero; 
 \item Partially massive model, $S_i$ basis, with \CP asymmetries fixed to zero and \qsq bins with widths half the size of those in the other configurations. 
 \end{enumerate}
Configuration (v) is included solely to enable comparison with Ref.~\cite{LHCb-PAPER-2020-002}. 

Pseudoexperiments are carried out to assess whether the extracted central values from the fit are unbiased, and whether the corresponding confidence intervals exhibit correct coverage. 
Observables with a relative bias or over/underestimation of the statistical uncertainty of more than 10\% are corrected using a classical Neyman construction to find the best estimates of the true central values and 68\% 
intervals~\cite{Neyman:1937uhy}.  Due to background contributions and the large number of fit parameters, the signal PDF can become negative in some regions of fit-variable phase space. To mitigate this unphysical effect, pseudodata for the Neyman construction are generated from a set of angular observables chosen to be as close as possible to those obtained from the data fit, while ensuring a positive PDF. The values used for the angular observables in the pseudoexperiment generation are updated for every point in the confidence belt for a given observable of interest. The central value is defined as the point in the confidence belt where the median most closely matches the fit result from data for the observable of interest. The positive PDFs are obtained with two methods: one varies the angular observables independently and tests whether the PDF is positive by sampling from the resulting angular model; the other varies the underlying decay amplitudes and subsequently infers the values of the angular observables from the result. 
Both methods use a random walk to minimize the Euclidean distance between the resulting positive PDF and the PDF obtained from the fit to data. The difference in the resulting intervals and biases from the two methods is assigned as a systematic uncertainty.
The results with and without these coverage and bias corrections are reported in the \textsc{HEPdata} page~\cite{hepdata.167733}.

The data are analyzed using two independent implementations of the PDF and data-simulation corrections from different control modes.
Both frameworks give compatible results when fitting pseudoexperiments and data, with residual discrepancies attributed to variations in the acceptance function, accounted for by systematic uncertainties, or to differences in the \decay{\Bp}{\Kp\mumu} veto. Given the good level of agreement between the two approaches, the baseline framework was chosen randomly. 

The fit results are found to have no significant dependence on run period, PID response, magnet polarity or the requirements made on the BDTs trained to reject the combinatorial and semileptonic backgrounds. The signal decay rate that results from the measured observables is also found to respect the symmetry relations given in Ref.~\cite{Alguero:2021yus}.
The same BDT goodness-of-fit method from the fit of the acceptance simulation is used, as well as a point-to-point dissimilarity method~\cite{Williams:2010vh}, to check that the fit to the data is of good quality. 

Systematic uncertainties on the observables are estimated using pseudoexperiments. A wide range of variations of the model used to generate these pseudoexperiments is considered, including the variation of relevant nuisance parameters, acceptance parameterisations, alternative fit model parameterisations, and injection of background events. 
The systematic uncertainties for all configurations are documented in the \textsc{HEPdata} page~\cite{hepdata.167733}.
The dominant systematic uncertainty depends on the observable and \qsq bin in question. For the P- and S-wave observables the systematic uncertainties are generally less than 25\% relative to the corresponding statistical uncertainties. 
At low \qsq the leading systematic uncertainties typically arise from the hardware trigger correction, from residual peaking background contamination and from the choice of polynomial orders in the angular background model.
For the S-wave and interference observables, the modelling of the S-wave lineshape in \mkpi gives rise to the dominant systematic uncertainty and is often similar to or larger than the statistical uncertainty. Effects arising from a $\Kp\pim$ system in D-wave and higher partial-wave states are confirmed to be negligible in the analyzed \mkpi region. For the relative P-wave branching fraction, the dominant uncertainty is due to the residual backgrounds for the normalisation mode.

The systematic uncertainties are evaluated coherently across all the \qsq bins. Therefore, the full covariance matrix of all observables across all bins is presented for use in subsequent phenomenological analyses. This is particularly pertinent for the large systematic uncertainty arising from the uncertainty on the branching fraction of the normalisation mode and, for the interference observables, the large systematic uncertainties arising from the choice of S-wave lineshape.

\begin{figure}[!tb]
    \centering
    \includegraphics[width=0.48\linewidth]{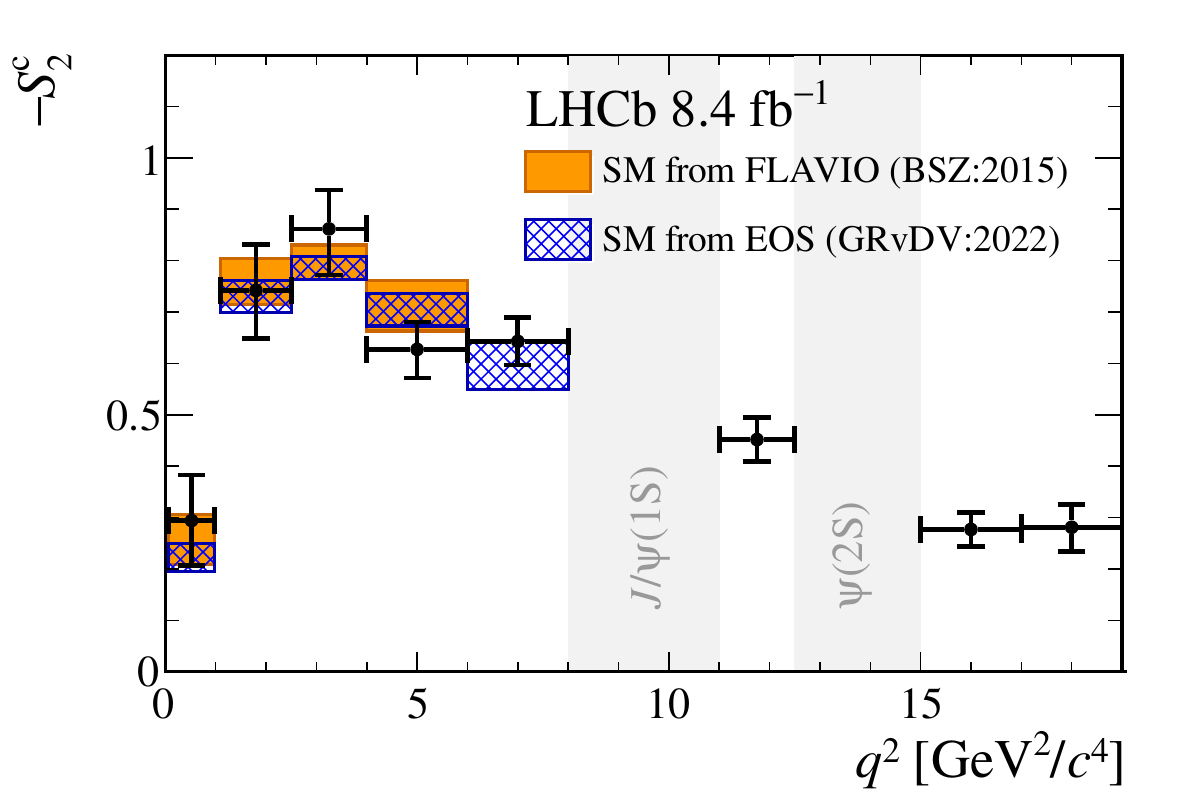} 
    \includegraphics[width=0.48\linewidth]{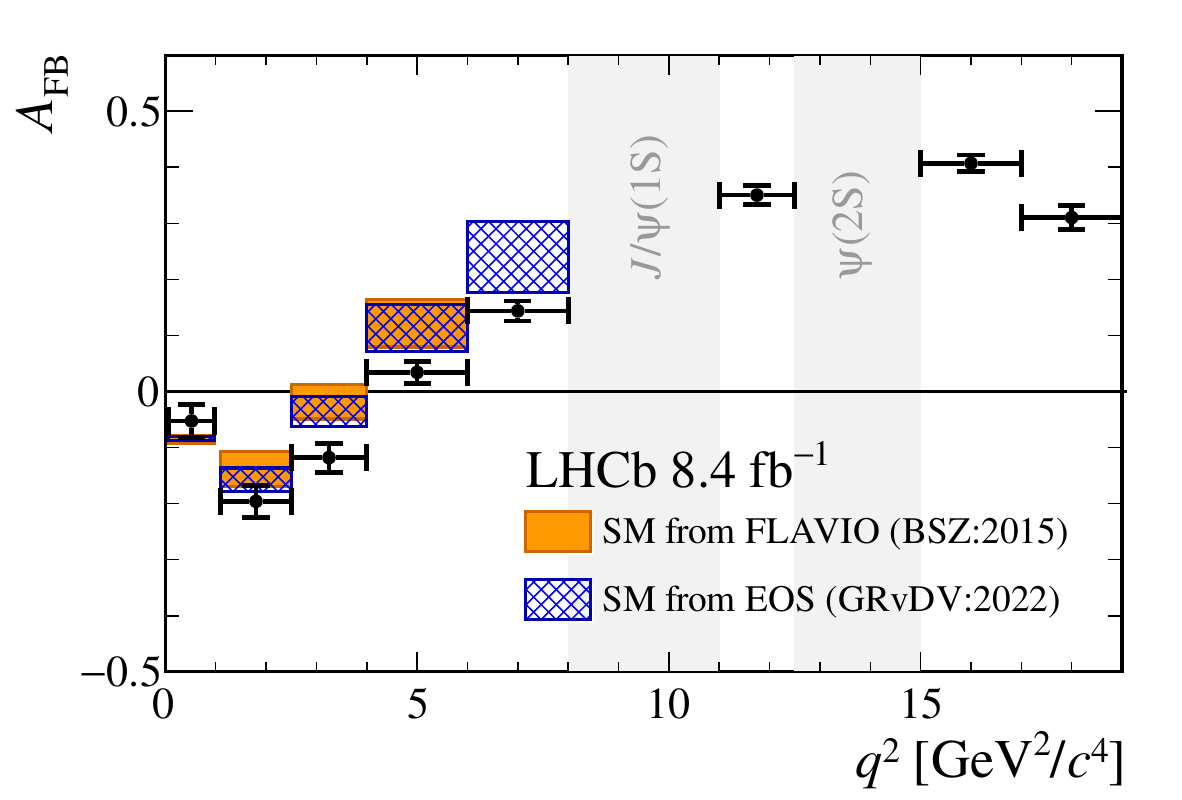} \\
    \includegraphics[width=0.48\linewidth]{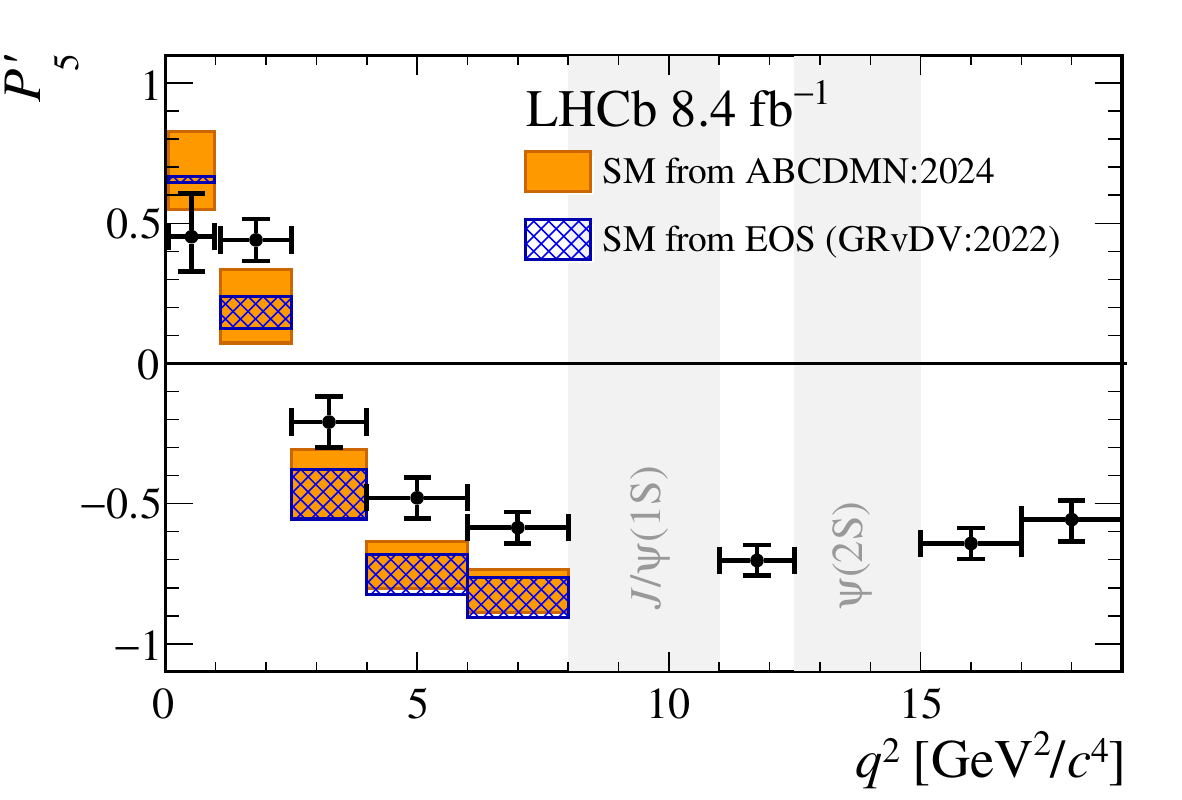} 
    \includegraphics[width=0.48\linewidth]{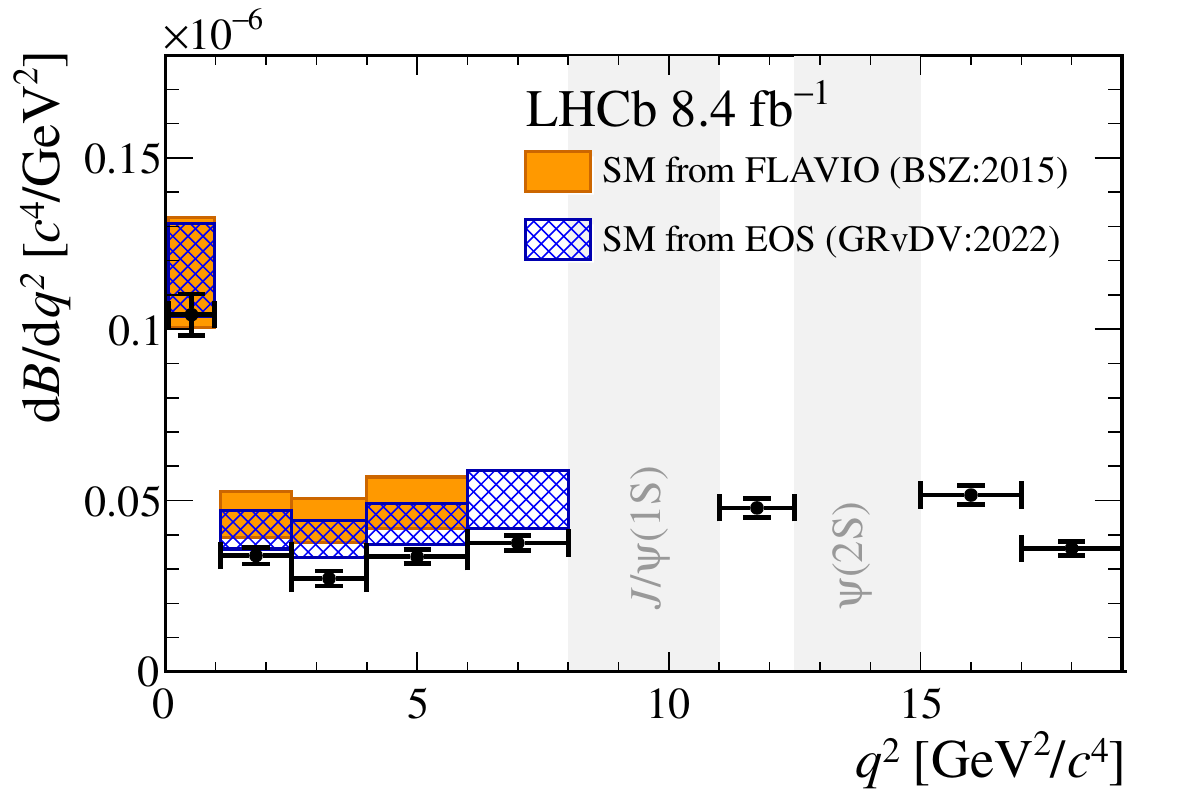} \\
    \caption{
    Results for the \CP-averaged angular observables $-S_{2}^c$, \AFB and $\deriv \BF /\deriv q^2$ from fit configuration (i), and $P'_5$ from fit configuration (ii), in bins of \qsq. In the approximation of massless leptons, $-S_{2}^c$ is equivalent to the longitudinal polarisation fraction $F_{\rm L}$.
    The data are compared to the SM predictions described in the text. The error bars show the total uncertainties.}
    \label{fig:results:angobs}
\end{figure}

The $\CP$-asymmetry observables all agree with SM expectations, with no significant $\CP$ asymmetry in the rate or angular observables. The $\CP$-averaged observables $S_{2}^c$, \AFB, $P'_5$ and $\deriv \BF/ \deriv q^2$ that are obtained from the fit configurations~(i)~and~(ii) are shown together with their respective SM predictions in Fig.~\ref{fig:results:angobs}. The optimized observable $P_{5}^{\prime}$ is defined as $P_{5}^{\prime} = \frac{S_{5}}{2\sqrt{-S_{2}^{c}S_{2}^{s}}}$, whilst in the approximation of massless leptons, the longitudinal polarisation fraction $F_{\rm L}$ is equivalent to $-S_2^c$. The results for all observables from all fit configurations are given in the \textsc{HEPdata} page~\cite{hepdata.167733}.
Different SM predictions are also shown, obtained using: the \flavio package~\cite{Straub:2018kue}, where local form factors and long-distance effects are parameterized according to Ref.~\cite{Bharucha:2015bzk} (BSZ) and Ref.~\cite{Altmannshofer:2014rta}, respectively; the EOS package~\cite{EOSAuthors:2021xpv}, where form factors are taken from Ref.~\cite{Gubernari_2022, Gubernari:2018wyi, Horgan:2015vla} (GRvDV); or directly from Ref.~\cite{Alguero:2023jeh} (ABCDMN).
Taking into account the overlap between the data samples, the results of the fit configuration~(v) are in excellent agreement with those reported in Ref.~\cite{LHCb-PAPER-2020-002}. 

Considering the observables individually, the results are largely in agreement with the SM predictions. However, tensions with SM predictions in localized regions of phase space have broadly increased relative to the levels reported in Ref.~\cite{LHCb-PAPER-2020-002}. The $P'_5$ observable in the \mbox{$4.0<q^2<6.0\gevgevcccc$} and \mbox{$6.0<q^2<8.0\gevgevcccc$} bins is discrepant with the the GRvDV (ABCDMN) predictions at $2.6$ and $2.7$ standard deviations~($\sigma$)~ ($2.1$ and $2.4\sigma$), respectively. Additionally, $P'_5$ is in tension with the same predictions in the \mbox{$1.1<q^2<2.5\gevgevcccc$} bin at the level of $2.7\sigma$ ($1.6\sigma$). The forward-backward asymmetry, \AFB, shows discrepancies with the BSZ (GRvDV) SM predictions in the \mbox{$2.5<q^2<4.0\gevgevcccc$} and \mbox{$4.0<q^2<6.0\gevgevcccc$} bins, with tension at the level of 2.5 and $1.9\sigma$ (2.2 and $1.7\sigma$). The differential branching fraction is also $2.1\sigma$ ($1.5\sigma$) below the BSZ (GRvDV) SM predictions when considered in the \qsq bin \mbox{$1.1<q^2<6.0\gevgevcccc$}.

The \flavio~\cite{Straub:2018kue} and EOS~\cite{EOSAuthors:2021xpv} software packages are used to fit the configuration~(i) results, varying the parameter $\Real(\C9 )$. For the \flavio package, the default values of the SM nuisances are used, including QCD effects as in Ref.~\cite{Bharucha:2015bzk,Altmannshofer:2014rta}, while for the EOS package,  values are taken from Ref.~\cite{Gubernari_2022}.
The $3.3\sigma$ discrepancy with respect to the SM value of $\Real(\C9 )$ obtained with \flavio in Ref.~\cite{LHCb-PAPER-2020-002} becomes $3.6\sigma$ ($3.8\sigma$) with \flavio (EOS). Including the differential branching fraction, the discrepancy increases to $4.1\sigma$ ($4.0\sigma$). Neglecting the $6.0<\qsq<8.0\gevgevcccc$ bin that is closest to the charmonium resonances reduces the observed significance of $\Real(\C9 )$ by about $0.2\sigma$.
The best fit to the angular distribution is obtained with a shift in the SM value of $\Real(\C9 )$ by $-0.94^{+0.21}_{-0.17}$ ($-0.92\pm0.23$) with \flavio (EOS).
When combining the information from the angular observables with the branching fraction, a shift in $\Real(\C9 )$ of $-0.93^{+0.18}_{-0.16}$ ($-0.94\pm0.22$) is obtained with \flavio (EOS).
Differences between these numbers and those reported in Refs.~\cite{LHCb-PAPER-2023-033, LHCb-PAPER-2024-011}, which fit for the Wilson coefficients directly, are due to the model inputs used, including the use of LHCb data to constrain the size of nonlocal contributions. 

In summary, using proton-proton collision data collected by the LHCb experiment corresponding to an integrated luminosity of $8.4{\mbox{\,fb}^{-1}}$, the rare flavor-changing neutral-current decay $\BdToKstmm$ has been analyzed. For the first time, this analysis reports the complete set of interference observables between the $\Kp\pim$ P- and S-wave contributions, accounts for lepton mass effects and makes measurements in narrow \qsq bins and determines the branching fraction without depending on a model of the angular distribution. The complete set of \CP asymmetries is determined and found to be consistent with zero. The results show a significant deviation from SM predictions~\cite{Bharucha:2015bzk,Altmannshofer:2014rta,Straub:2018kue,EOSAuthors:2021xpv} for the \CP-averaged observables.

\section*{Acknowledgements}
%
%
\noindent We express our gratitude to our colleagues in the CERN
accelerator departments for the excellent performance of the LHC. We
thank the technical and administrative staff at the LHCb
institutes.
We acknowledge support from CERN and from the national agencies:
ARC (Australia);
CAPES, CNPq, FAPERJ and FINEP (Brazil); 
MOST and NSFC (China); 
CNRS/IN2P3 (France); 
BMFTR, DFG and MPG (Germany);
INFN (Italy); 
NWO (Netherlands); 
MNiSW and NCN (Poland); 
MCID/IFA (Romania); 
MICIU and AEI (Spain);
SNSF and SER (Switzerland); 
NASU (Ukraine); 
STFC (United Kingdom); 
DOE NP and NSF (USA).
We acknowledge the computing resources that are provided by ARDC (Australia), 
CBPF (Brazil),
CERN, 
IHEP and LZU (China),
IN2P3 (France), 
KIT and DESY (Germany), 
INFN (Italy), 
SURF (Netherlands),
Polish WLCG (Poland),
IFIN-HH (Romania), 
PIC (Spain), CSCS (Switzerland), 
and GridPP (United Kingdom).
We are indebted to the communities behind the multiple open-source
software packages on which we depend.
Individual groups or members have received support from
Key Research Program of Frontier Sciences of CAS, CAS PIFI, CAS CCEPP, 
Minciencias (Colombia);
EPLANET, Marie Sk\l{}odowska-Curie Actions, ERC and NextGenerationEU (European Union);
A*MIDEX, ANR, IPhU and Labex P2IO, and R\'{e}gion Auvergne-Rh\^{o}ne-Alpes (France);
Alexander-von-Humboldt Foundation (Germany);
ICSC (Italy); 
Severo Ochoa and Mar\'ia de Maeztu Units of Excellence, GVA, XuntaGal, GENCAT, InTalent-Inditex and Prog.~Atracci\'on Talento CM (Spain);
SRC (Sweden);
the Leverhulme Trust, the Royal Society and UKRI (United Kingdom).

\clearpage 
\section*{End Matter}
\subsection*{The \texorpdfstring{$\bm{{\Bz\to\Kp\pim\mumu}}$}{B0->K*mumu} differential decay rate}
\label{sup:decay_rate}
The decay may proceed via P- or S-wave amplitudes (higher partial waves are negligible in the \mkpi region analyzed). The angular basis is described in Ref.~\cite{LHCb-PAPER-2013-019}. The differential decay rate, described in Ref.~\cite{Alguero:2021yus}, for \Bzb (\Bz) is
{\footnotesize
\begin{align}
\label{supp:eq:pdf}
&\frac{1}{\deriv(\Gamma + \overline{\Gamma})/\deriv q^2} \bigg|_{\text{P+S}} \frac{\deriv^5\GammaorGammabar}{\deriv q^2 \deriv m_{K\pi} \deriv\cos\theta_\ell \deriv\cos\theta_K \deriv\phi}\nonumber\\
&= (1 - \hat{F_{\rm S}}) \frac{9}{64\pi} \Bigg[
\zeta_{1}^{s} \sin^2\theta_K + \zeta_{1}^{c} \cos^2\theta_K + \zeta_{2}^{s} \sin^2\theta_K\cos2\theta_{\ell} + \zeta_{2}^{c} \cos^2\theta_K \cos 2\theta_\ell \nonumber \\
& + \zeta_{3} \sin^2\theta_K \sin^2\theta_\ell \cos 2\phi + \zeta_{4} \sin 2\theta_K \sin 2\theta_\ell \cos\phi + \zeta_{5} \sin 2\theta_K \sin\theta_\ell \cos\phi \nonumber\\
& + \zeta_{6}^{s} \sin^2\theta_K\cos\theta_{\ell} + \zeta_{6}^{c} \cos^2\theta_K \cos\theta_\ell + \zeta_{7} \sin 2\theta_K \sin\theta_\ell \sin\phi \nonumber\\
& + \zeta_{8} \sin 2\theta_K \sin 2\theta_\ell \sin\phi + \zeta_{9} \sin^2\theta_K \sin^2\theta_\ell \sin 2\phi 
\Bigg] \times |\mathcal{BW}_{\text{P}}(m_{K\pi})|^2 \nonumber\\
&+ \frac{1}{8\pi} \Bigg[ \left( \tilde{\zeta}_{1a}^{c} + \tilde{\zeta}_{2a}^{c} \cos 2\theta_\ell \right) \times |\mathcal{L}_{\text{S}}(m_{K\pi})|^2 \nonumber\\
&+ \frac{1}{2} \cos\theta_K \left( \tilde{\zeta}_{1b}^{c,\, {\rm re}} \Real[\mathcal{L}_{\text{S}}(m_{K\pi}) \mathcal{BW}_{\text{P}}^*(m_{K\pi})] - \tilde{\zeta}_{1b}^{c,\,{\rm im}}\Imag[\mathcal{L}_{\text{S}}(m_{K\pi}) \mathcal{BW}_{\text{P}}^*(m_{K\pi})] \right) \nonumber\\
&+ \frac{1}{2} \cos\theta_K \cos 2\theta_\ell \left( \tilde{\zeta}_{\rm S1}^{\rm re}  \Real[\mathcal{L}_{\text{S}}(m_{K\pi}) \mathcal{BW}_{\text{P}}^*(m_{K\pi})] - \tilde{\zeta}_{\rm S1}^{\rm im}  \Imag[\mathcal{L}_{\text{S}}(m_{K\pi}) \mathcal{BW}_{\text{P}}^*(m_{K\pi})] \right) \nonumber\\
&+ \frac{1}{2} \sin\theta_K \sin 2\theta_\ell \cos\phi \left( \tilde{\zeta}_{\rm S2}^{\rm re} \Real[\mathcal{L}_{\text{S}}(m_{K\pi}) \mathcal{BW}_{\text{P}}^*(m_{K\pi})] - \tilde{\zeta}_{\rm S2}^{\rm im}  \Imag[\mathcal{L}_{\text{S}}(m_{K\pi}) \mathcal{BW}_{\text{P}}^*(m_{K\pi})] \right) \nonumber\\
&+ \frac{1}{2} \sin\theta_K \sin\theta_\ell \cos\phi \left( \tilde{\zeta}_{\rm S3}^{\rm re}  \Real[\mathcal{L}_{\text{S}}(m_{K\pi}) \mathcal{BW}_{\text{P}}^*(m_{K\pi})] - \tilde{\zeta}_{\rm S3}^{\rm im}  \Imag[\mathcal{L}_{\text{S}}(m_{K\pi}) \mathcal{BW}_{\text{P}}^*(m_{K\pi})] \right) \nonumber\\
&+ \frac{1}{2} \sin\theta_K \sin\theta_\ell \sin\phi \left( \tilde{\zeta}_{\rm S4}^{\rm re}\Imag[\mathcal{L}_{\text{S}}(m_{K\pi}) \mathcal{BW}_{\text{P}}^*(m_{K\pi})] + \tilde{\zeta}_{\rm S4}^{\rm im }\Real[\mathcal{L}_{\text{S}}(m_{K\pi}) \mathcal{BW}_{\text{P}}^*(m_{K\pi})] \right) \nonumber\\
&+ \frac{1}{2} \sin\theta_K \sin 2\theta_\ell \sin\phi \left( \tilde{\zeta}_{\rm S5}^{\rm re}\Imag[\mathcal{L}_{\text{S}}(m_{K\pi}) \mathcal{BW}_{\text{P}}^*(m_{K\pi})] + \tilde{\zeta}_{\rm S5}^{\rm im }\Real[\mathcal{L}_{\text{S}}(m_{K\pi}) \mathcal{BW}_{\text{P}}^*(m_{K\pi})] \right) 
\Bigg],
\end{align}
\noindent }where $\zeta_{i} \equiv S_{i} \mp A_{i}$ for $\Bz$ (\Bzb) decays.
The functions $\mathcal{BW}_{\rm P}(m_{K\pi})$ and $\mathcal{L}_{\rm S}(m_{K\pi})$ describe the $\mkpi\equiv m_{K\pi}$ dependence of the P- and S-wave amplitudes, respectively. 
The observables $\zeta_{i}$ are normalized by the P-wave rate, whilst those labeled $\tilde{\zeta}_{i}$ are normalized by the sum of P- and S-wave rates.
The relative S-wave fraction is denoted as \mbox{$\hat{F}_{\rm S} = \frac{\Gamma_{\rm S} + \overline{\Gamma}_{\rm S}}{\Gamma_{\rm P} + \overline{\Gamma}_{\rm P} + \Gamma_{\rm S} + \overline{\Gamma}_{\rm S}}$} $= 2\tilde{S}_{1a}^{c} - \tfrac{2}{3}\tilde{S}_{2a}^{c}$. 
The observable $F_{\rm S}$ is defined as $\tilde{S}_{2a}^{c} \equiv -\tfrac{3}{8}F_{\rm S}$, such that in the massless lepton limit one finds $\hat{F}_{\rm S} = F_{\rm S}$.  
The coefficient $S_{6}^{s}$ is related to the forward-backward asymmetry through $S_{6}^{s} = \tfrac{4}{3}A_{\rm FB} - \tfrac{1}{2}S_{6}^{c}$, as discussed in Ref.~\cite{Altmannshofer:2008dz}.

Due to the choice of measuring rate-normalized observables, one of the $S_{i}$ terms must be removed from the fit via 
\begin{equation}
    1 = \tfrac{3}{4}\left( 2S_{1}^{s} + S_{1}^{c} \right) - \tfrac{1}{4}\left(2S_{2}^{s} + S^{c}_{2} \right).
    \end{equation}
\renewcommand{\theequation}{R.\arabic{equation}}
\setcounter{equation}{0}
The following additional relations between observables are used to simplify fitting models:
\begin{equation}
\label{eq:noscalar}
 S_{6}^{c} = 0, \qquad A_{6}^{c} = 0;
\end{equation}
\begin{equation}
\label{eq:pwavemassless2}
    S_{1}^{s} = 3S_{2}^{s}, \qquad A_{1}^{s} = 3A_{2}^{s};
\end{equation}
\begin{equation}
\label{eq:swavemassless}
    \tilde{S}_{2a}^{c} = -\tilde{S}_{1a}^{c}, \quad 
    \tilde{A}_{1a}^{c} = -\tilde{A}_{2a}^{c}, \quad 
    \tilde{S}_{2b}^{c,\,\rm re/im} = -\tilde{S}_{1b}^{c,\,\rm re/im}, \quad 
    \tilde{A}_{2b}^{c,\,\rm re/im} = -\tilde{A}_{1b}^{c,\,\rm re/im};
\end{equation}
\begin{equation}
\label{eq:pwavemassless1}
    S_{1}^{c} = -S_{2}^{c}, \qquad A_{1}^{c} = -A_{2}^{c}.
\end{equation}
Equation \ref{eq:noscalar} holds if there are no new physics scalar or tensor amplitudes. Furthermore, if in addition the ratio $\frac{4m^2_{\mu}}{\qsq}$ is zero, then \ref{eq:pwavemassless2}-\ref{eq:pwavemassless1} hold. The massless model (configurations (iv) and (v)) assumes all the relations \ref{eq:noscalar}-\ref{eq:pwavemassless1}, the partially massive model (configurations (i), (ii) and (vi)) assumes relations~\ref{eq:noscalar}-\ref{eq:swavemassless}, and the fully massive model (configuration (iii)) assumes none of these relations. Pseudoexperiment studies show the relation~\ref{eq:pwavemassless1} causes larger biases on the angular observables than~\ref{eq:pwavemassless2}. This is likely due to the dependence of $S_{1}^{c}$ on the time-like amplitude. Dropping the assumption~\ref{eq:pwavemassless1}, as is done for the partially massive model, reduces this bias to negligible levels for $q^{2}$ values above $\sim1\gevgevcccc$, below which the fully massive model is used. In other \qsq regions, using the fully massive model results in a loss of sensitivity significantly larger than the small bias introduced when using the partially massive model.

\renewcommand{\theequation}{\arabic{equation}}

\subsection*{An illustration of the \texorpdfstring{$\bm{P_5^{\prime}}$}{P5'} observable}
\begin{figure}[h!]
    \centering
    \includegraphics[width = 0.6\textwidth]{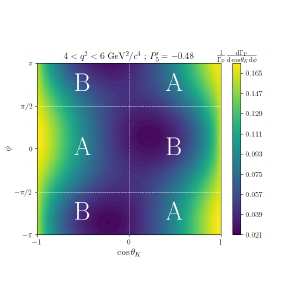}
    \caption{\label{fig:sfive}The angular distribution of the $P_{5}^{\prime}$ term of the angular decay rate projected onto the $\phi$ and $\cos\theta_{K}$ angular variables. The plot is split into four quadrants. While found from a full angular fit, the value of $P_5'$ is proportional to the asymmetry of yields between regions $A$ and $B$.}
\end{figure}

 The observable $P_5'$ (or $S_5$) is the coefficient of the angular function $\sin 2\theta_K\,\sin\theta_\ell\,\cos\phi$ in the equation that describes the differential decay rate. 
The projection of the fitted P-wave differential decay rate on  the $\phi$--$\cos\theta_K$ plane is shown in Fig.~\ref{fig:sfive}. While the value of $P_5'$ is found from a full angular fit, its value is proportional to the relative asymmetry in the signal yields between the regions labeled as A and B in the plot.

\clearpage




\addcontentsline{toc}{section}{References}
\bibliographystyle{LHCb}
\bibliography{main,standard,LHCb-PAPER,LHCb-CONF,LHCb-DP,LHCb-TDR}

\newpage
\centerline
{\large\bf LHCb collaboration}
\begin
{flushleft}
\small
R.~Aaij$^{38}$\lhcborcid{0000-0003-0533-1952},
A.S.W.~Abdelmotteleb$^{58}$\lhcborcid{0000-0001-7905-0542},
C.~Abellan~Beteta$^{52}$\lhcborcid{0009-0009-0869-6798},
F.~Abudin\'en$^{58}$\lhcborcid{0000-0002-6737-3528},
T.~Ackernley$^{62}$\lhcborcid{0000-0002-5951-3498},
A.A.~Adefisoye$^{70}$\lhcborcid{0000-0003-2448-1550},
B.~Adeva$^{48}$\lhcborcid{0000-0001-9756-3712},
M.~Adinolfi$^{56}$\lhcborcid{0000-0002-1326-1264},
P.~Adlarson$^{86}$\lhcborcid{0000-0001-6280-3851},
C.~Agapopoulou$^{14}$\lhcborcid{0000-0002-2368-0147}
% ...

{\footnotesize \it

% ...
$^{14}$Universit{\'e} Paris-Saclay, CNRS/IN2P3, IJCLab, Orsay, France\\
% ...
$^{38}$Nikhef National Institute for Subatomic Physics, Amsterdam, Netherlands\\
% ...
$^{48}$Instituto Galego de F{\'\i}sica de Altas Enerx{\'\i}as (IGFAE), Universidade de Santiago de Compostela, Santiago de Compostela, Spain\\
% ...
$^{52}$Physik-Institut, Universit{\"a}t Z{\"u}rich, Z{\"u}rich, Switzerland\\
% ...
$^{56}$H.H. Wills Physics Laboratory, University of Bristol, Bristol, United Kingdom\\
% ...
$^{58}$Department of Physics, University of Warwick, Coventry, United Kingdom\\
% ...
$^{61}$School of Physics and Astronomy, University of Glasgow, Glasgow, United Kingdom\\
$^{62}$Oliver Lodge Laboratory, University of Liverpool, Liverpool, United Kingdom\\
% ...
$^{70}$Syracuse University, Syracuse, NY, United States\\
% ...
$^{86}$Department of Physics and Astronomy, Uppsala University, Uppsala, Sweden, associated to $^{61}$\\
% ...
}
\end{flushleft}

\end{document}